\documentclass[aip,preprint]{revtex4-1}
\usepackage{graphicx}
\usepackage[section]{placeins}
 \usepackage{float}

\begin{document}


\title{Parametrized path integral formulation for large fermion systems}

\author{Yunuo Xiong}

\affiliation{College of Science, Zhejiang University of Technology, Hangzhou 31023, China}

\author{Hongwei Xiong}
\email{xionghw@zjut.edu.cn}


\affiliation{College of Science, Zhejiang University of Technology, Hangzhou 31023, China}


\begin{abstract}
The exchange antisymmetry between identical fermions gives rise to the well known fermion sign problem, in the form of large cancellation between positive and negative contribution to the partition function, making any simulation methods which directly sample this partition function exponentially difficult to converge. In this work, we employ path integral molecular dynamics (PIMD) and build upon the recently discovered fictitious particle model to investigate the fermion sign problem further. We consider the validity and invalidity condition for the method of parametrized path integral formulation of the partition function and extrapolation to circumvent the fermion sign problem. For the valid region of our method, our simulation shows that we may give accurate prediction of the energy for large fermion systems, which is much beyond the capability of the direct sampling in the traditional method. In particular, we find and verify a simple universal relation for high temperature noninteracting particles or strongly repulsive interacting particles at low temperatures.
\end{abstract}


\maketitle

\section{Introduction}

Recently, a polynomial time algorithm \cite{Hirshberg} for path integral molecular dynamics (PIMD) simulations for identical bosons has been found, by utilizing a recursive formula to calculate the bosonic partition function in polynomial time. The PIMD technique has been successfully applied to identical bosons to predict supersolid phase in high pressure Deuterium \cite{Deuterium}. However, like its path integral Monte Carlo (PIMC) counterpart \cite{CeperRMP,boninsegni1,boninsegni2,Dornheim,DornheimMod}, for large fermion systems at low temperatures, PIMD suffers from the infamous fermion sign problem \cite{ceperley,Alex,troyer,loh,lyubartsev,vozn,Science,Wu,Umrigar,Li,Wei,Yao2} as well.

Very recently, a fictitious particle model \cite{XiongFSP} inspired by the recursive formula of PIMD \cite{HirshbergFermi,Xiong,Xiong2,Xiong3,Xiong4,Xiong5} has been proposed. The fictitious particle model is characterized by a real parameter $\xi$ which provides continuous interpolation between bosons ($\xi=1$), distinguishable particles ($\xi=0$) and fermions ($\xi=-1$). Previously, an attempt was made to circumvent the fermion sign problem by running simulations for $\xi\geq 0$, which does not suffer from fermion sign problem, and extrapolate the thermodynamic properties to fermions ($\xi=-1$). It is expected that this method would be an alternative to  restricted path integral Monte Carlo \cite{nodes,Helium,Militzer}, density matrix quantum Monte Carlo \cite{Blunt,Malone}, configuration PIMC \cite{Schoof1, Schoof2,Schoof3, Yilmaz}, permutation blocking PIMC \cite{PB1,PB2}, and auxiliary field quantum Monte Carlo \cite{Joonho} for fermion sign problem at finite temperatures. However, our approach relies on whether the extrapolation is sensible and there is no simple way \textit{a priori} to judge. Hence, in our previous work, the extrapolation method was applied only for a few fermions.

In this work, we give systematic studies on the validity and invalidity condition of the extrapolation with our method of parametrized path integral molecular dynamics. We find by numerical simulation and general physical analysis that for large number of noninteracting fermions, with increasing temperature, $E(\xi)$ in the whole region $-1\leq\xi\leq 1$ may be a monotonic convex function or concave function, which provides the basis for the accuracy of our method. For noninteracting fermion systems with an inflection point near $\xi=0$, however, the extrapolation is no longer valid anymore. In particular, with increasing repulsive interaction strength, we notice the gradual disappearance of the inflection point in the noninteracting particles at temperature far below the Fermi temperature, and so the extrapolation becomes accurate again. The present work provides an efficient and accurate method to circumvent the fermion sign problem for large noninteracting or interacting fermion systems in polynomial time. The analysis of the parametrized partition function also leads to a simple universal relation for high temperature noninteracting particles or strongly interacting particles at low temperatures, which has been long overlooked.

\section{Fermion sign problem and parametrized partition function}

\subsection{Fermion sign problem}

We consider the following Hamiltonian for $N$ particles
\begin{equation}
\hat H=\frac{1}{2m}\sum_{l=1}^N\hat \textbf p_l^2+\hat V(\textbf r_1,\cdots,\textbf r_N).
\end{equation}
Here $V(\textbf r_1,\cdots,\textbf r_N)$ includes both the external potential and inter-particle interactions. 
The partition function of $N$  particles is
\begin{equation}
Z(\beta)=Tr(e^{-\beta \hat H}).
\end{equation}
Here $\beta=1/k_B T$, with $k_B$ being the Boltzmann constant and $T$ being the system temperature.
The average energy is
\begin{equation}
E(\beta)=-\frac{\partial\ln Z(\beta)}{\partial \beta}.
\label{energyZ}
\end{equation}

In coordinate representation to consider the trace in the partition function for identical bosons and fermions, we have
\begin{equation}
Z_{F/B}(\beta)=\sum_{p\in S_N}(\pm 1)^{N_p}\int d\textbf{r}_1d\textbf{r}_2\cdots d\textbf{r}_N\left<p\{\textbf{r}\}|e^{-\beta \hat H}|\{\textbf{r}\}\right>.
\label{Xipartition}
\end{equation}
Here $\{\textbf{r}\}$ denotes $\{\textbf{r}_1,\cdots,\textbf{r}_N\}$.  $S_N$ represents the set of $N!$ permutation operations denoted by $p$. The factor $(\pm 1)^{N_p}$ is due to the exchange effect of identical particles, with $N_p$ a number defined to be the minimum number of times for which pairs of indices must be interchanged in permutation $p$ to recover the original order. For example, two interchanges must be applied on the permutation $(2,1,4,3)$ to recover the original order $(1,2,3,4)$ and so for this particular permutation $N_p$ is equal to 2.  In $\pm 1$ of the above expression, we have $+1$ for identical bosons ($Z_B$), while $-1$ for identical fermions ($Z_F$). 

Using $e^{-\beta\hat H}=e^{-\Delta\beta\hat H}\cdots e^{-\Delta\beta\hat H}$ with $\Delta\beta=\beta/P$ and the technique of path integral \cite{feynman,kleinert,Tuckerman}, the partition function $Z_{F/B}(\beta)$ can be mapped as a classical system of interacting ring polymers \cite{chandler,Parrinello,Miura,Cao,Cao2,Jang2,Ram,Kinugawa,Roy,Roy1,Roy2,Roy3,Kinugawa1,Kinugawa2,Roy4,Roy5,Roy6,Poly,Craig,Braa,Haber,Thomas}. 
In recent works by Hirshberg et al. \cite{Hirshberg,HirshbergFermi}, a recursion formula is found to calculate the partition function for both bosons ($\xi=1$) and fermions ($\xi=-1$). Based on the path integral ring polymer, the fermion partition function ($Z_F$) and boson partition function ($Z_B$) are expressed as an integral for an analytical function of the coordinates of interacting ring polymers 
\begin{equation}
Z_{F/B}\sim\int d\textbf{R}_1\cdots d\textbf{R}_N e^{-\beta U_{F/B}^{(N)}}.
\end{equation}
Here the integrand is a function of a set of ring polymer coordinates $(\textbf{R}_1,\cdots,\textbf{R}_N)$, with $\textbf{R}_i=(\textbf{r}_i^1,\cdots,\textbf{r}_i^P)$ corresponding to $P$ ring polymer coordinates for the $i$th particle. 
A recursion formula is found to give the expression of $U_F^{(N)}$ for fermions and $U_B^{(N)}$ for bosons. For bosons, $U_B^{(N)}$ is always real and $e^{-\beta U_{B}^{(N)}}$ may be sampled for large boson systems \cite{Hirshberg,Deuterium,Xiong,Xiong2}. However, because of the factor $(-1)^{N_p}$ for fermions, $U_F^{(N)}$ is a complex function, which can not be sampled with available methods. 

To deal with the situation of complex function $U_F^{(N)}$, in the usual traditional method  to deal with the fermion sign problem, the fermion partition function may be rewritten as
\begin{equation}
Z_{F}\sim\int d\textbf{R}_1\cdots d\textbf{R}_N e^{-\beta U_{B}^{(N)}}e^{\beta(U_B^{(N)}-U_F^{(N)})}.
\end{equation}
In this case, we may still use molecular dynamics to sample $e^{-\beta U_{B}^{(N)}}$, while the energy is 
\begin{equation}
E(\beta)=\frac{\left<\epsilon s\right>_B}{{\left<s\right>_B}}.
\label{Energys}
\end{equation}
Here $s=e^{\beta(U_B^{(N)}-U_F^{(N)})}$ and $\epsilon$ is the estimator for energy. The average $\left<\cdots\right>_B$ is about the samples based on the partition function $Z_B$ for bosons. Unfortunately, this simple and elegant method can not be applied to large fermion systems at low temperatures because of the fermion sign problem.

At low temperatures below the Fermi temperature, roughly speaking, we have
\begin{equation}
\left<s\right>_B=e^{-\beta (F_F-F_B)} \sim A e^{-\beta(E_F-E_B)}.
\label{Sequation}
\end{equation}
Here $F_F$ and $F_B$ are the free energy of fermions and bosons, respectively. $E_F$ and $E_B$ are the average energy for fermions and bosons, respectively. We find for the examples in this work that $A\sim 10$. For $\left<s\right>_B\sim 10^{-3}$ or smaller, the traditional method to calculate the energy of fermions is no longer feasible \cite{Dornheim}. Because of the factor $\left<s\right>_B$ in Eq. (\ref{Energys}), the error in $\left<s\right>_B$ will cause the result to be unreliable since the fluctuations of $s$ would be larger or even much larger than $\left<s\right>_B$. Because of the small factor $\left<s\right>_B$ for large fermion systems at low temperature, it is hard to calculate the energy, both in path integral Monte Carlo \cite{ceperley,Alex,Science,CeperRMP,boninsegni1,boninsegni2,Dornheim,DornheimMod} and path integral molecular dynamics \cite{HirshbergFermi,Xiong2}. General consideration shows that the simulation becomes exponentially hard to converge with increasing numbers of fermions and decreasing temperatures \cite{ceperley}. This is known as the infamous fermion sign problem \cite{ceperley,Alex,loh,troyer,lyubartsev,vozn,Science,Wu,Umrigar,Li,Wei,Yao2}. The present work will show that for the situation of $\left<s\right>_B<< 10^{-3}$, we still have hope to predict accurately and efficiently the energy of large fermion systems.

\subsection{Parametrized partition function to circumvent the fermion sign problem}

We consider the following parametrized partition function with a real parameter $\xi$:
\begin{equation}
Z(\beta,\xi)\sim\sum_{p\in S_N}\xi^{N_p}\int d\textbf{r}_1d\textbf{r}_2\cdots d\textbf{r}_N\left<p\{\textbf{r}\}|e^{-\beta \hat H}|\{\textbf{r}\}\right>.
\label{Xipartition}
\end{equation}
In our previous work \cite{XiongFSP}, using $e^{-\beta\hat H}=e^{-\Delta\beta\hat H}\cdots e^{-\Delta\beta\hat H}$ and the technique of path integral, we have shown that there is still a recursion formula for the parameterized partition function $Z(\beta,\xi)$. Here we only give the result of this recursion formula,

\begin{equation}
Z(\beta,\xi)\sim\int d\textbf{R}_1\cdots d\textbf{R}_N e^{-\beta U_\xi^{(N)}}.
\label{Zxi}
\end{equation}
$U_\xi^{(N)}$ is given by
\begin{equation}
U_\xi^{(N)}=-\frac{1}{\beta}\ln W_\xi^{(N)}+\frac{1}{P}\sum_{j=1}^P V\left(\textbf{r}_1^j,\cdots,\textbf{r}_N^j\right),
\label{UFN}
\end{equation}
and $W_\xi^{(N)}$ is
\begin{equation}
W_\xi^{(N)}=\frac{1}{N}\sum_{k=1}^N\xi^{k-1}e^{-\beta E_N^{(k)}}W_\xi^{(N-k)}.
\label{WFN}
\end{equation}
In addition,
\begin{equation}
E_N^{(k)}=\frac{1}{2}m\omega_P^2\sum_{l=N-k+1}^N\sum_{j=1}^P\left(\textbf r_l^{j+1}-\textbf{r}_l^j\right)^2.
\end{equation}
Here $\textbf r_l^{P+1}=\textbf r_{l+1}^1$, except for $l=N$ for which $\textbf r_N^{P+1}=\textbf r_{N-k+1}^1$.
In addition, $\omega_P=\sqrt{P}/\beta\hbar$.

From Eq. (\ref{energyZ}), it is straightforward to get the energy estimator (see Ref. \cite{XiongFSP} for details), so that once we get the importance sampling of $e^{-\beta U_\xi^{(N)}}$ for the parametrized partition function by molecular dynamics, we can obtain the average energy $E(\beta,\xi)$ of the system with any parameter $\xi\geq 0$.

In Ref. \cite{XiongFSP}, we propose a method to circumvent the fermion sign problem by an extrapolation with polynomial function to fit the data $E(\beta,\xi\geq 0)$. In that work, however, the range of applicability of the extrapolation method is not clear, which limits the practical use of the method for large fermion systems. Fortunately, the present work will solve this problem, which gives reasonable prediction for large fermion systems much beyond the capability of the traditional method. The analysis of the parametrized partition function also leads to a simple universal relation for high temperature noninteracting particles or strongly repulsive interacting particles at low temperatures.

\section{General behavior of $E(\xi)$ for large noninteracting fermion systems and the validity condition of the extrapolation} 
\label{ideal}

\begin{figure}[htbp]
\begin{center}
 \includegraphics[width=0.9\textwidth]{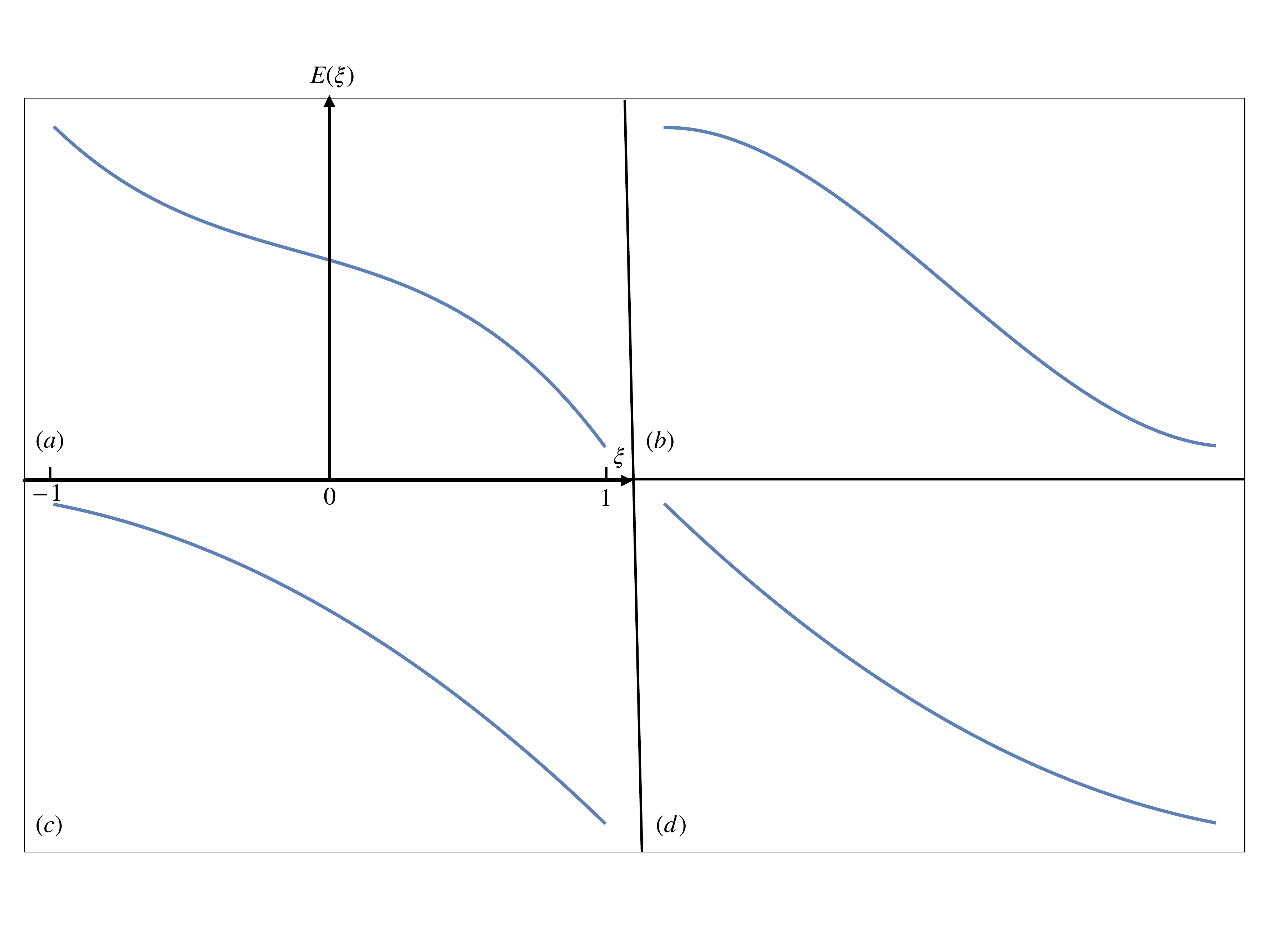} 
\caption{\label{qualitative} When the monotonic behavior of $E(\xi)$ is considered, shown are four typical cases for $E(\xi)$. In (a) and (b), there is an inflection point at $\xi=0$, while in (c) and (d), $E(\xi)$ has good analytical property without the inflection point.}
\end{center}
\end{figure}

Taking into account the monotonic behavior \cite{XiongFSP} of $E(\xi)$, in principle, we have different possibilities of the behavior of  $E(\xi)$ shown in Fig. \ref{qualitative}. For the situations of Fig. \ref{qualitative}(a) and Fig. \ref{qualitative}(b), there is an inflection point at $\xi=0$. In this case, it is hard to predict accurately the value of $E(\xi=-1)$ by the extrapolation based on the data of $E(\xi)$ for positive $\xi$. For the situations of Fig. \ref{qualitative}(c) and Fig. \ref{qualitative}(d) without the inflection point, we have good chance to predict accurately the value of $E(\xi=-1)$ by the extrapolation with the fitting of the data of $\xi\geq 0$. In this work, the so-called good analytical property of $E(\xi)$ means that: 

(i) $E(\xi)$ has monotonic behavior; 

(ii) $E(\xi)$ has no inflection point between $-1\leq \xi\leq 1$. 

We will verify by numerical experiments and general physical analysis that increasing temperature or increasing repulsive interaction strength will contribute to the good analytical property of $E(\xi)$.

For large number of particles, it is well known that the energy is almost the same for canonical ensemble and grand canonical ensemble. This suggests that for the parametrized partition function $Z(\xi,\beta)$, the energy $E(\xi,\beta)$ for noninteracting particles may be calculated by the following formula in grand canonical ensemble:
\begin{equation}
N=\sum_{\textbf n}\frac{1}{e^{\beta({\epsilon({\textbf n})-\mu})}-\xi},
\label{chemical}
\end{equation}

\begin{equation}
E(\xi,\beta,N)=\sum_{\textbf n}\frac{\epsilon(\textbf n)}{e^{\beta({\epsilon({\textbf n})-\mu})}-\xi}.
\label{energyc}
\end{equation}
$\epsilon({\textbf n})$ is the single-particle eigenenergies of the system. With given parameters of $N,\beta$, and $\xi$,  from Eq. (\ref{chemical}), we can get the chemical potential $\mu(\xi,\beta,N)$. Using further Eq. (\ref{energyc}), we can get the energy $E(\xi,\beta,N)$ in the grand canonical ensemble.

\begin{figure}[htbp]
\begin{center}
 \includegraphics[width=0.9\textwidth]{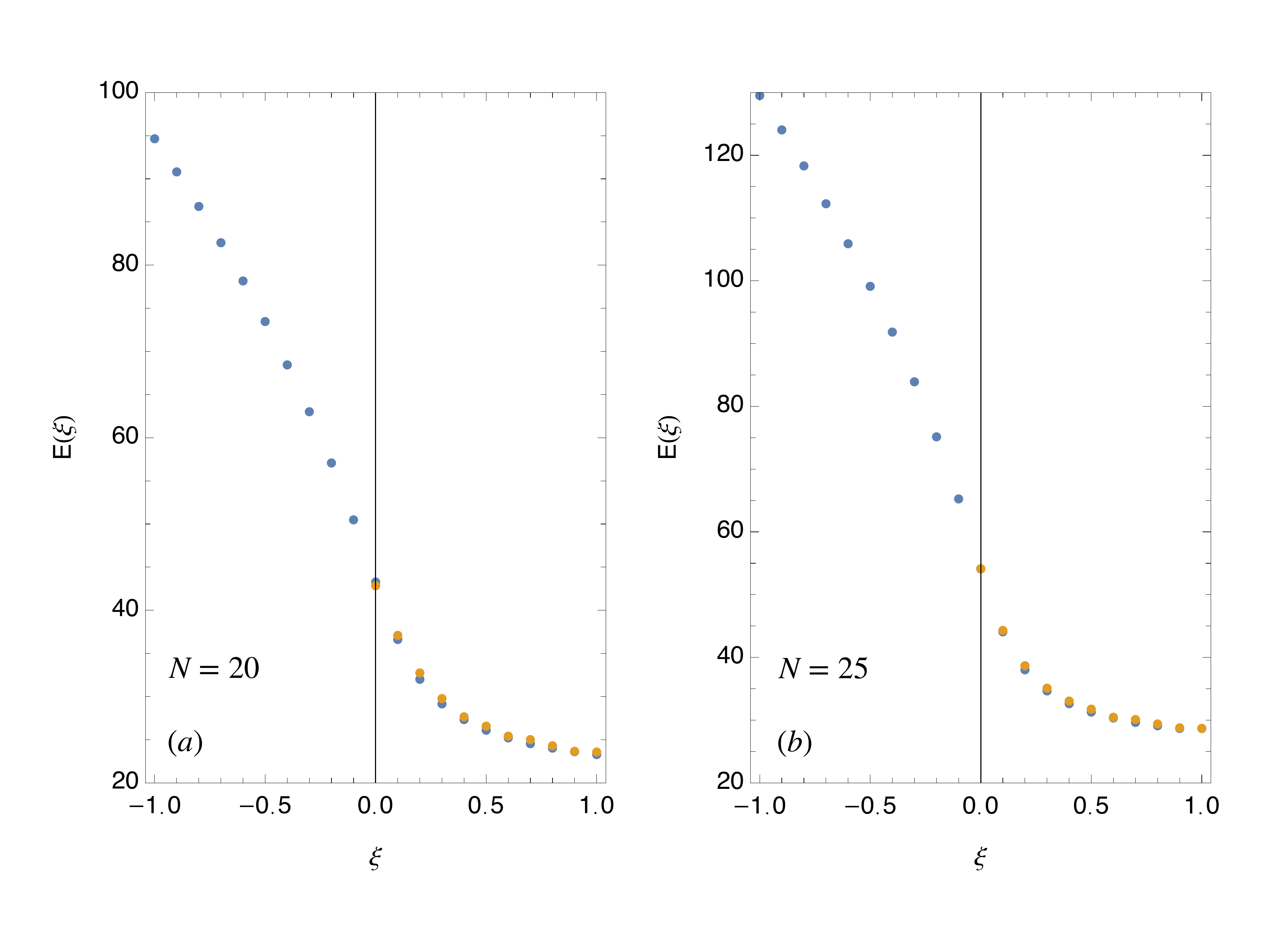} 
\caption{\label{ComparisonPIMD} (a) For $N=20$ and $\beta=1$, the blue circle shows $E(\xi)$ based on Eqs. (\ref{chemical}) and (\ref{energyc}), while the orange circle for $\xi\geq 0$ is the result of PIMD. (b) shows the results for $N=25$ and $\beta=1$. Perfect agreement is found for these two different methods.}
\end{center}
\end{figure}

For a two-dimensional harmonic trap $\frac{1}{2}m\omega^2(x^2+y^2)$ with the unit $\hbar=m=1$ and $\omega=1$, in Fig. \ref{ComparisonPIMD}, we show the energy by PIMD and the result based on the above two equations (\ref{chemical}) and (\ref{energyc}) for 20 and 25 particles, respectively. We do find perfect agreement for $\xi\geq 0$ for large number of particles with $N=20$ and $N=25$ for $\beta=1$ in our calculations. This suggests that Eqs. (\ref{chemical}) and (\ref{energyc}) can be applied to the whole region of $-1\leq \xi\leq 1$. In Fig. \ref{ComparisonPIMD}, we also give the energy for $\xi\leq 0$ based on Eqs. (\ref{chemical}) and (\ref{energyc}).

\begin{figure}[htbp]
\begin{center}
 \includegraphics[width=0.9\textwidth]{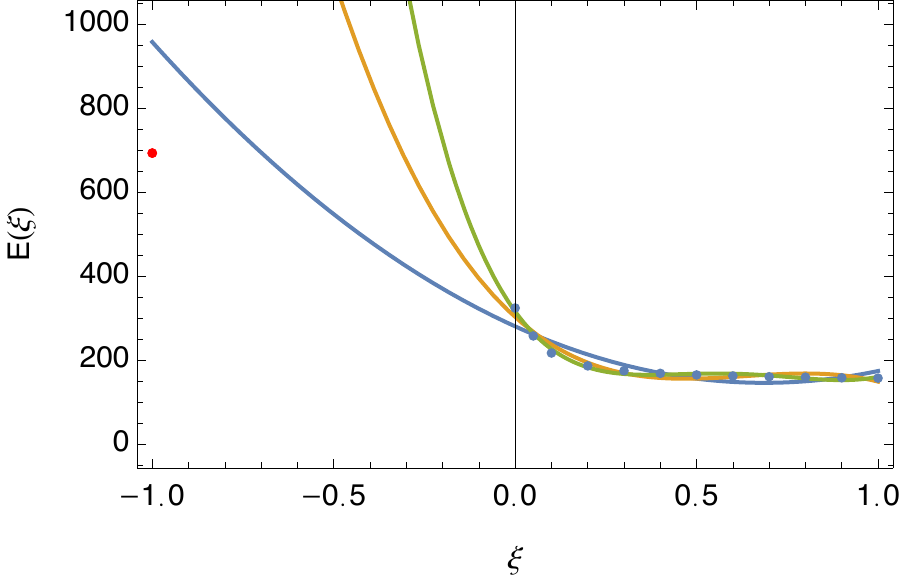} 
\caption{\label{100ideal} For $N=100$ and $\beta=10$, shown is the fitting of the data by PIMD with different fitting function. The blue line is for $\sum_{j=0}^2 a_j\xi^j$, the orange line for $\sum_{j=0}^3 a_j\xi^j$, and the green line for $\sum_{j=0}^4 a_j\xi^j$. The red circle is the energy of fermions in the grand canonical ensemble. All these fitting functions fail to predict the correct value of the energy of fermions.}
\end{center}
\end{figure}

For both examples shown in Fig. \ref{ComparisonPIMD}, we notice an inflection point at $\xi\approx 0$. If we use the parabolic function to fit the data of $E(\xi<0)$, we find that $\frac{d^2E}{d\xi^2}<0$, while $\frac{d^2E}{d\xi^2}>0$ for the fitting of the data between $0\leq\xi\leq 1$. In the presence of the inflection point at $\xi\approx 0$, by extrapolation based on the data $E(\xi\geq 0)$, it would be hard to predict the correct value for $E(\xi=-1)$. For 100 noninteracting particles with $\beta=1$ as an example, in Fig. \ref{100ideal}, we show the failure of various polynomial functions for fitting and extrapolation to predict the value of $E(\xi=-1)$. 
Fortunately, we will show in due course that for many other examples, even the simple parabolic function for fitting can give accurate prediction. This suggests that our method can be applied for some situations, while it becomes invalid for other cases. The question is to determine the range of application of our extrapolation method based on the parametrized partition function.

\begin{figure}[htbp]
\begin{center}
 \includegraphics[width=0.7\textwidth]{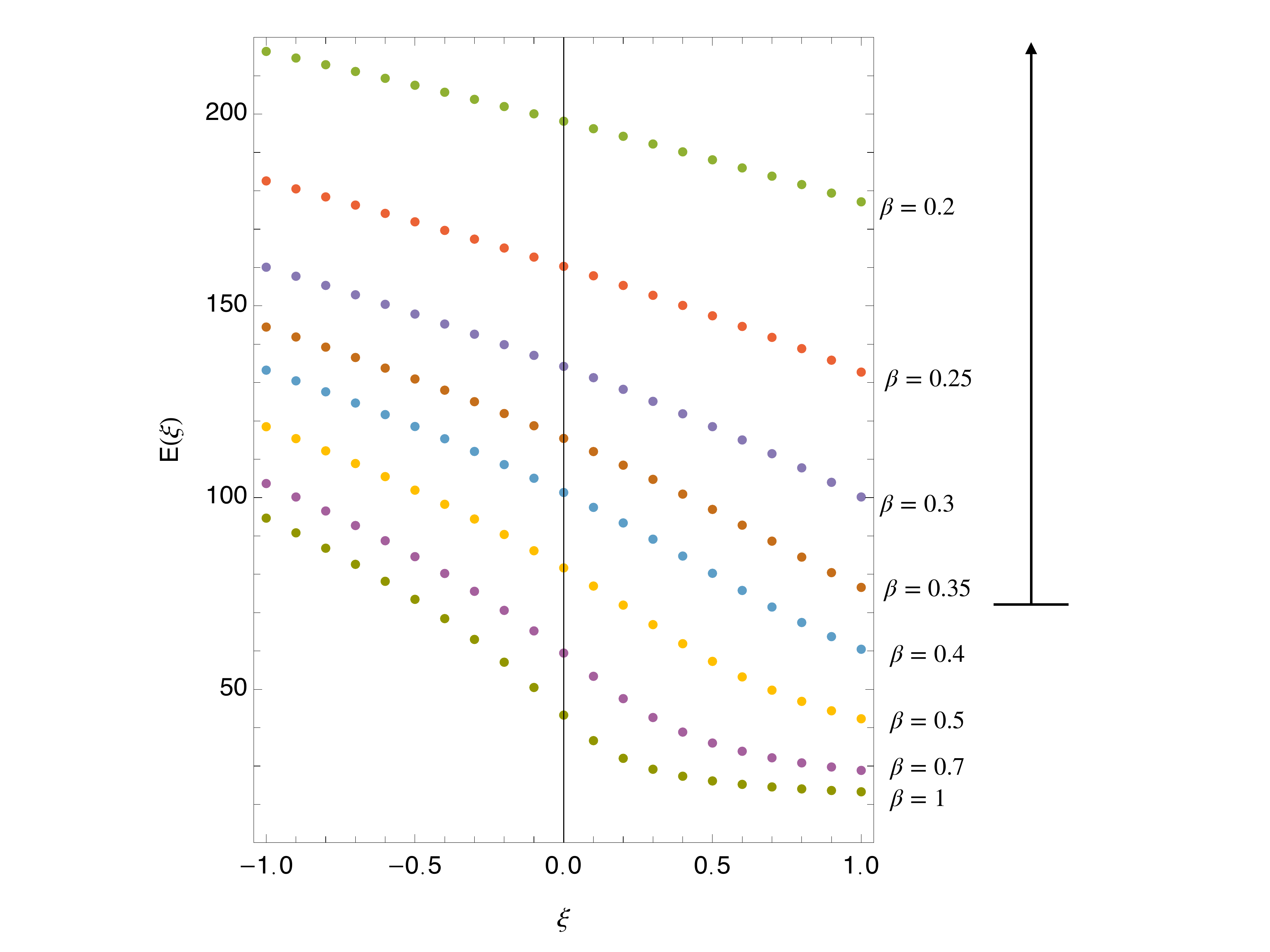} 
\caption{\label{20idealbeta} Shown is $E(\xi)$ for 20 noninteracting particles in the two-dimensional harmonic trap with different $\beta$. We notice the disappearance of the inflection point for sufficiently small $\beta$. It is clearly shown that $E(\xi)$ has good analytical property for $\beta\leq 0.35$.}
\end{center}
\end{figure}

To answer the above question, we consider numerical experiments for 
20 noninteracting particles with different temperatures. The Fermi temperature is $T_F=6.08$. By increasing the temperature (decreasing $\beta$), based on the calculations by Eqs. (\ref{chemical}) and (\ref{energyc}), we notice a gradual disappearance of the inflection point, as shown in Fig. \ref{20idealbeta}. One of the physical reasons lies in that in the limit of infinite temperature, $E(\xi)$ is a constant. Hence, we expect that the behavior of $E(\xi)$ will become simpler as the temperature increases. We will give a more careful explanation later. At the Fermi temperature, we have $\beta\approx 0.16$. Hence, in Fig. \ref{20idealbeta}, what we consider is the low temperature situation below the Fermi temperature, where the quantum statistics is important, and we encounter nasty fermion sign problem. 

\begin{figure}[htbp]
\begin{center}
 \includegraphics[width=0.7\textwidth]{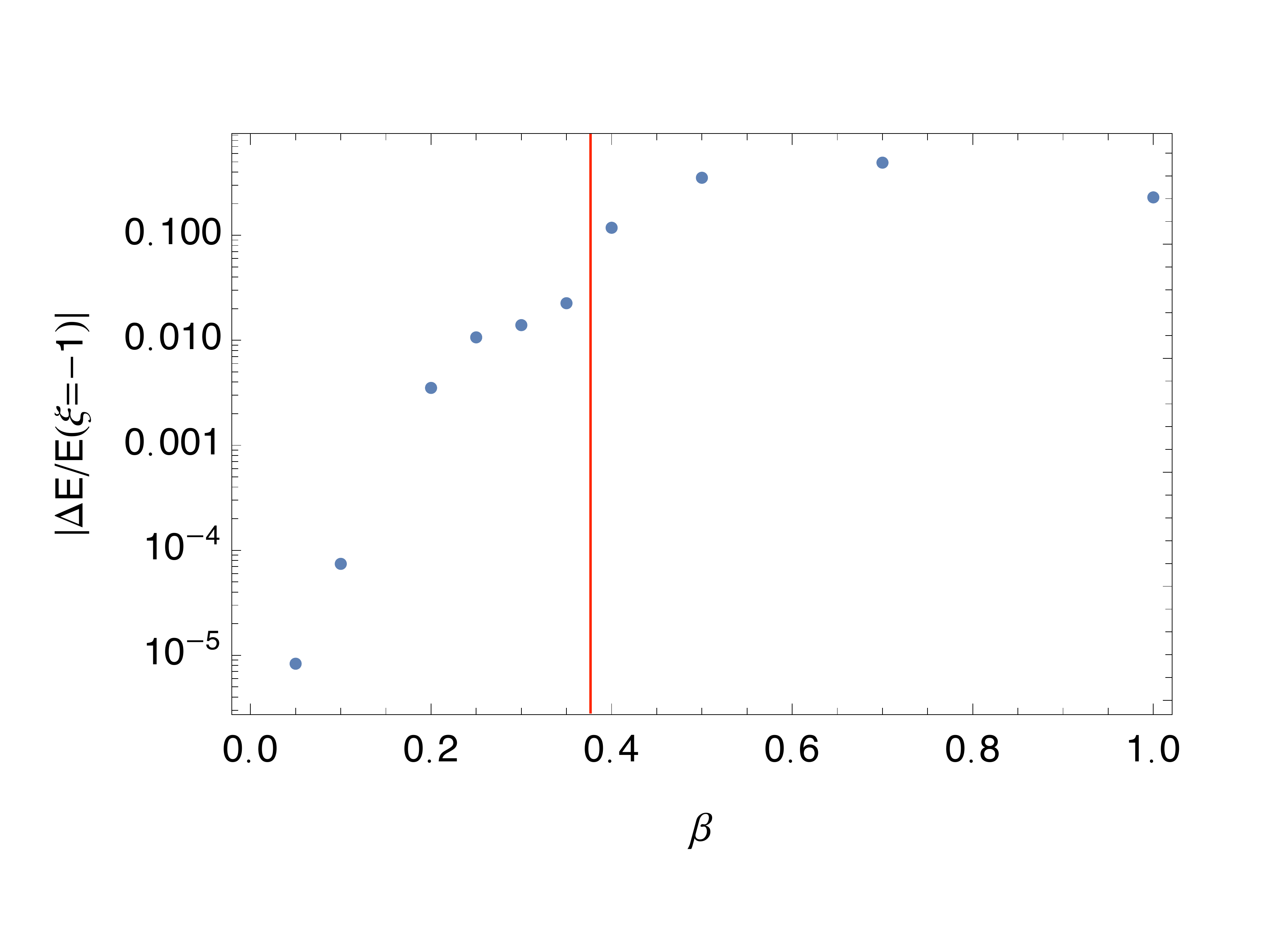} 
\caption{\label{20idealerror} The $\log_{10}$ plot of $|\Delta E/E(\xi=-1)|$ for different $\beta$. The red line denotes a sharp boundary for the validity and invalidity region of the parametrized PIMD. $\Delta E$ is the difference between the prediction of the parametrized PIMD and the energy of fermions in the grand canonical ensemble.}
\end{center}
\end{figure}

As shown in Fig. \ref{20idealbeta}, for 20 noninteracting particles with different $\beta$, we see that for $\beta\leq 0.35$, $E(\xi)$ is a concave function in the whole region $-1\leq\xi\leq 1$, while for $\beta\geq 0.5$ there exists an obvious inflection point. Although it is hard to judge with the naked eye for $\beta=0.4$, the fitting for $E(\xi\leq 0)$ and $E(\xi\geq 0)$ separately show that $E(\xi)$ is not a good analytical function for this situation. This gradual disappearance of the inflection point is due to the change of $E(0\leq\xi\leq 1)$ from convex function to a concave function for $\beta\leq 0.35$, while $E(-1\leq\xi\leq 0)$ is always a convex function. Hence, it is safe to judge the validity of the method of parametrized PIMD by considering whether $E(0\leq\xi\leq 1)$ by PIMD is a concave function for noninteracting situation. We do find that for $\beta\leq 0.35$ without the inflection point, the simple parabolic function can already give very good prediction for $E(\xi=-1)$. For $\beta\geq 0.4$, however, it seems that we can not give reasonable prediction with any polynomial function for fitting and extrapolation. In Fig. \ref{20idealerror}, we show $log_{10}|\Delta E|/E(\xi=-1)$ for different $\beta$ by the fitting with parabolic function and extrapolation. $\Delta E$ is the difference between the prediction of the parametrized PIMD and the energy of fermions in grand canonical ensemble.

\begin{figure}[htbp]
\begin{center}
 \includegraphics[width=0.7\textwidth]{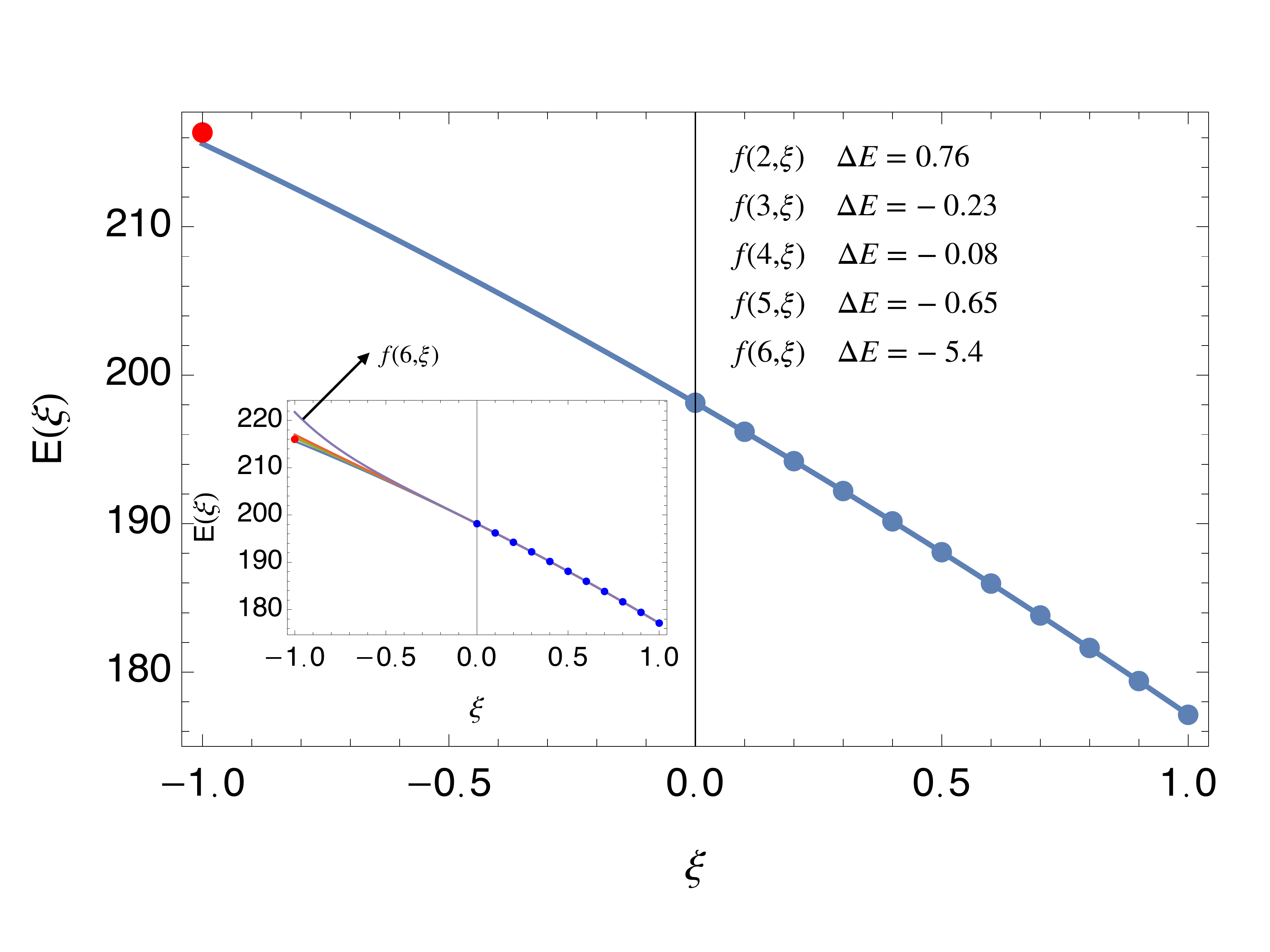} 
\caption{\label{20beta02} Shown is the fitting for $N=20$ and $\beta=0.2$ with the parabolic function. The red circle is the energy of the fermions in the grand canonical ensemble. The inset in the lower left corner shows the different fitting function up to $n=6$ with $f(n,\xi)=\sum_{j=1}^n a_j\xi^j$. We see that for $2\leq n\leq 5$, there is no obvious difference for different fitting function $f(n,\xi)$, while $f(6,\xi)$ gives obvious difference which changes the convex property of $E(\xi)$. This shows clearly that $f(6,\xi)$ leads to the overfitting of the data. The inset in the upper right corner gives the deviation from the energy in the grand canonical ensemble for different fitting function $f(n,\xi)$. For this example, $\left<s\right>_B\sim 0.005$ based on the estimation of Eq. (\ref{Sequation}), which means the possibility to get $E_F$ in the traditional method. In the method of parametrized partition function, we can calculate $E_F$ much more accurately and efficiently. As a comparison, for $N=20$ and $\beta=0.3$, $\left<s\right>_B\sim 0.00001$. The traditional method can not be used for this situation, while our method can still predict accurate value of $E_F$, as shown in Fig. \ref{20idealerror}.}
\end{center}
\end{figure}

Even for $\beta\leq 0.35$ where the parabolic function has already given good prediction, one may wonder what happens if general polynomial function is used for fitting and extrapolation. We consider the following polynomial function for the fitting of $E(\xi\geq 0)$:
\begin{equation}
f(n,\xi)=\sum_{j=0}^n a_j\xi^j.
\end{equation}
In Fig. \ref{20beta02}, for $N=20$ and $\beta=0.2$, we use the data of $E(\xi\geq 0)$ to predict the energy for fermions with parabolic function, which agrees well with the expected result. In the inset of this figure, we show the result 
with different fitting function $f(n,\xi)$. We see that for $2\leq n\leq 5$, we can give good prediction of the energy for fermions, while $f(6,\xi)$ gives the wrong answer. The wrong answer with $f(6,\xi)$ is not surprising because it is well known that too many parameters in a fitting with no regularization will always lead to overfitting.

Of course, we need a general method to avoid the overfitting. The inset in Fig. \ref{20beta02}  shows clearly that $f(6,\xi)$ is an overfitting because it deviates significantly from other fitting results and does not agree with the expected result. Even when we do not know the energy for fermions in advance, we have several methods to judge whether there is an overfitting with increasing $n$ for the fitting function $f(n,\xi)$. For  the situation where the parametrized PIMD can be applied, i.e. $E(\xi)$ is a monotonic convex or monotonic concave function between $-1\leq\xi\leq 1$,  we expect that the following situation is a clear signature of the overfitting for $f(n,\xi)$:

(i) $f(n,\xi)$ is no longer a monotonic function.

(ii) $f(n,\xi)$ is no longer a convex or concave function between $-1\leq\xi\leq 1$.

(iii) We expect that reasonable $f(n,\xi)$ should be a perturbation expansion for $\xi$. Hence, we may request that $|a_{j+1}|<|a_j|$ for $j\geq 2$. If this condition is not satisfied, it is a signature of overfitting.

(iv) If the predictions of $f(n-2,\xi)$ and $f(n-1,\xi)$ agree with each other, while $f(n,\xi)$ deviates significantly from the predictions of $f(n-2,\xi)$ and $f(n-1,\xi)$, it is quite possible that $f(n,\xi)$ is an overfitting.

The case of underfitting with too few parameters can be judged easier than overfitting, by the signature that $f(n,\xi)$ can not give good fitting for the data of $E(\xi\geq 0)$.

For the example shown in Fig. \ref{20beta02}, different fitting functions $f(n,\xi)$ are
\begin{equation}
f(2,\xi)=198.134-19.227 \xi-1.77086 \xi^2,
\end{equation}
\begin{equation}
f(3,\xi)=198.145-19.4087 \xi-1.29417 \xi^2-0.317793 \xi^3,
\end{equation}
\begin{equation}
f(4,\xi)=198.145-19.3913 \xi-1.38159 \xi^2-0.177933 \xi^3-0.0699301 \xi^4,
\end{equation}
\begin{equation}
f(5,\xi)=198.145-19.4018\xi -1.29465 \xi^2-0.422858\xi^3+0.210519 \xi^4-0.112179\xi^5,
\end{equation}
\begin{equation}
f(6,\xi)=198.145-19.4183 \xi-1.0924\xi^2-1.29273\xi^3+1.89137\xi^4-1.6073 \xi^5+0.49837\xi^6.
\end{equation}

Even without knowing other fitting result and the energy of fermions in advance, $f(6,\xi)$ does not satisfy the requirement of the perturbation expansion, which means that it is an overfitting. It is also not a concave function between $-1\leq\xi\leq 1$. This example shows that we have self-consistent method to judge whether there is an overfitting, which makes our method safe to predict the energy for fermions. In this example, $f(1, \xi)$ is an obvious underfitting.

For a question like the prediction of $E(\xi=-1)$ for $N=20,\beta=0.32$ by the method of parametrized partition function, we think it is best to think about more than just this question. We should consider a series of different $\beta$ from high temperature to low temperature, so that we have good sense of the range of application of our method, so that we have solid conclusion whether our method can be applied to the designated question. 

\begin{figure}[htbp]
\begin{center}
 \includegraphics[width=0.7\textwidth]{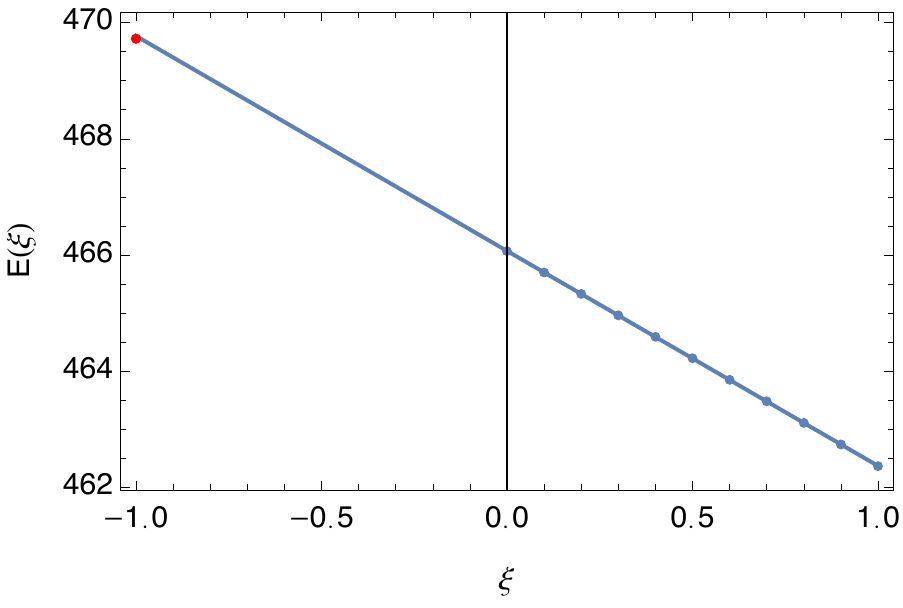} 
\caption{\label{20linear} For $N=20$ and $\beta=0.05$, the solid line is the linear fitting of the data of circles, and the red solid is the energy of fermions in the grand canonical ensemble.}
\end{center}
\end{figure}

As the temperature increases, we notice that the absolute value of the curvature of $E(\xi)$ decreases. 
For $T>>T_F$ with $T_F$ the Fermi temperature, the curvature of $E(\xi)$ approaches zero. In this case,  we expect the following simple and universal relation:
\begin{equation}
E(\xi=-1,\beta)-E(\xi=0,\beta)\simeq E(\xi=0,\beta)-E(\xi=1,\beta).
\label{relation}
\end{equation}
More specifically, we have the above relation if we find that $E(0\leq \xi\leq 1)$ satisfies well the linear relation. For example, for $N=20$ and $\beta=0.05$ shown in Fig. \ref{20linear}, a linear fitting gives only the error of $\Delta E=-0.048$, while $E(\xi=-1)-E(\xi=1)=7.35$. This means that there would be perfect linear relation even if there is significant energy difference between fermions and bosons with significantly different quantum statistics effect. The unusual characteristic of the above relation lies in that it holds even for $|a_1|>>|a_2|$, while in the usual consideration that the quantum statistics is not important for $T>>T_F$, the relation is
\begin{equation}
E(\xi=-1,\beta)\gtrsim E(\xi=0,\beta)\gtrsim E(\xi=1,\beta)
\end{equation}
for $T>>T_F$. Our relation (\ref{relation}) is much more refined, compared with the above famous expression. We will give the physical reason for the linear relationship (\ref{relation}) later.

\section{Large fermion systems with repulsive interaction}

In Sec. \ref{ideal}, we see that by increasing the temperature, $E(\xi,\beta)$ can have good analytical property so that the method of parametrized path integral partition function can be applied well. In this section, we will show by numerical experiments that the repulsive interactions have the effect of making $E(\xi,\beta)$ have good analytical property too, so that we have good chance to apply the method of parametrized path integral partition function to the situation of repulsive interaction.

We consider in this section the two-dimensional harmonic trap by including a Coulomb-type interaction:
\begin{equation} 
V_{int}=\sum_{l<j}^N \frac{\lambda}{{|{\textbf r}_l-{\textbf r}_j|}}.
\end{equation}
Here $\lambda$ represents the dimensionless coupling constant of the Coulomb-type interaction.

\begin{figure}[htbp]
\begin{center}
 \includegraphics[width=0.7\textwidth]{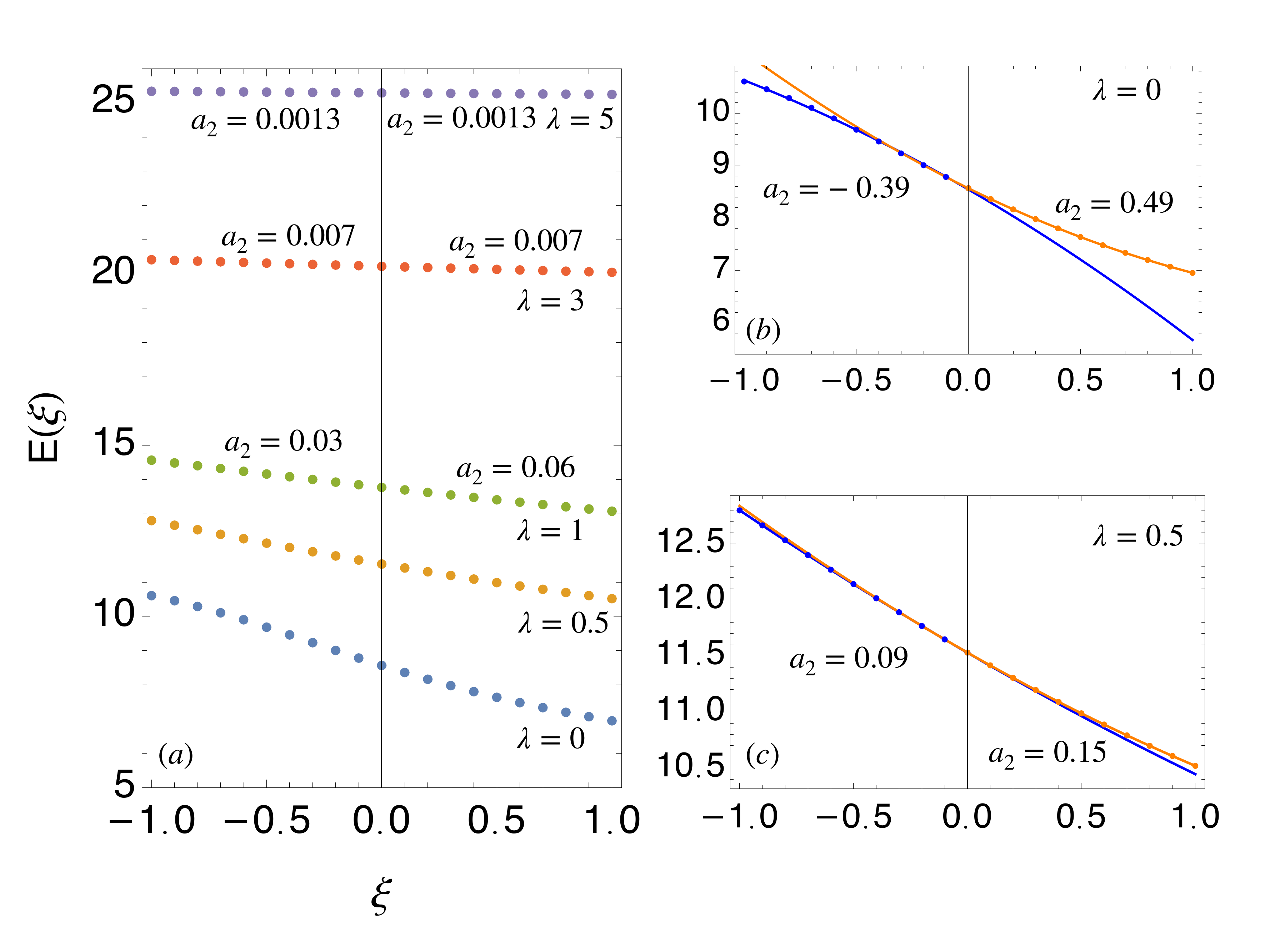} 
\caption{\label{4intfull} For 4 particles with $\beta=1$, (a) shows the energy $E(\xi)$ by the traditional method to deal with the parametrized partition function for different $\lambda$. (b) shows $a_2$ in the parabolic fitting for the data $E(\xi\leq 0)$ and the data $E(\xi\geq 0)$, respectively, for $\lambda=0$. (c) shows $a_2$ in the parabolic fitting for the data $E(\xi\leq 0)$ and the data $E(\xi\geq 0)$, respectively, for $\lambda=0.5$. We also give $a_2$ for $\lambda=1,3,5$ in (a) for the data on the left and the data on the right, respectively. We see that for $\lambda\geq 0.5$, $E(\xi)$ has good analytical property.}
\end{center}
\end{figure}

For $N=4,\beta=1$, we use the traditional method to calculate the energy $E(\xi)$ for different coupling constant $\lambda$. Fig. \ref{4intfull} clearly shows that as $\lambda$ increases, $E(\xi)$ can have good analytical property. With further increasing of $\lambda$, $E(\xi)$ satisfies good linear property.

\begin{figure}[htbp]
\begin{center}
 \includegraphics[width=0.7\textwidth]{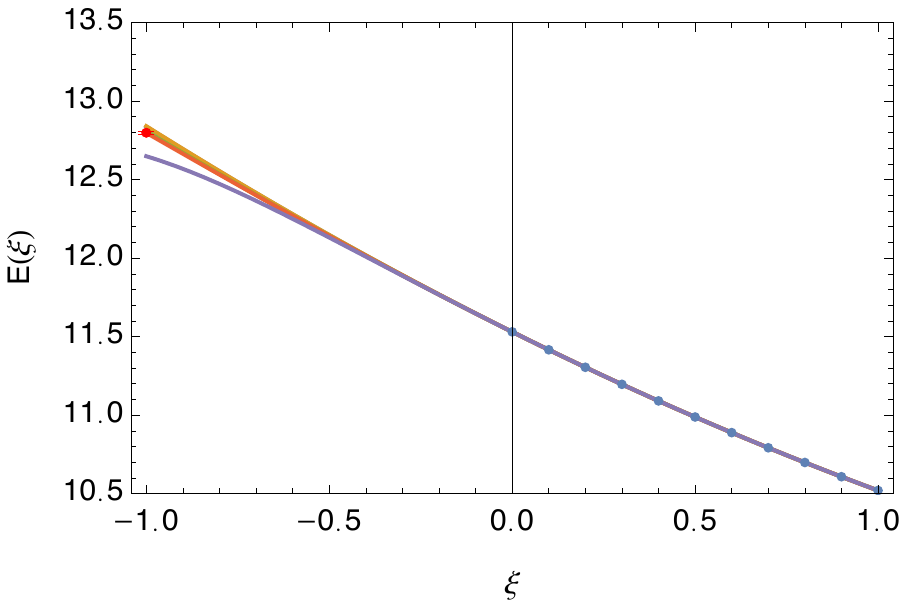} 
\caption{\label{4int05} For $N=4,\beta=1$, and $\lambda=0.5$, the fitting functions of $f(2,\xi)$, $f(3,\xi)$, $f(4,\xi)$, and $f(5,\xi)$ can give good prediction, while $f(6,\xi)$ leads to the overfitting. The red circle is the result of the traditional method. $\left<s\right>_B=0.25$ in the traditional method, while $\left<s\right>_B\sim 0.5$ based on the estimation of Eq. (\ref{Sequation}).}
\end{center}
\end{figure}

For sufficiently large $\lambda$ so that $E(\xi)$ has good analytical property, it is expected that the data of $E(\xi\geq 0)$ can predict the correct energy for fermions.
As an example, we consider the situation of $\lambda=0.5$ for 4 particles shown in Fig. \ref{4int05}.
The fitting functions of Fig. \ref{4int05} are
\begin{equation}
f(2,\xi)=11.5302-1.15874\xi+0.148427\xi^2,
\end{equation}
\begin{equation}
f(3,\xi)=11.5302-1.15923\xi+0.149702\xi^2-0.0008497\xi^3,
\end{equation}
\begin{equation}
f(4,\xi)=11.5301-1.15737\xi+0.140411\xi^2+0.0140161\xi^3-0.00743288\xi^4,
\end{equation}
\begin{equation}
f(5,\xi)=11.5301-1.15724\xi+0.139319\xi^2+0.0170919\xi^3-0.0109548\xi^4+0.00140878\xi^5,
\end{equation}
\begin{equation}
f(6,\xi)=11.5301-1.1567+0.132995\xi^2+0.04429\xi^3-0.0635\xi^4+0.0482\xi^5-0.0156\xi^6.
\end{equation}
The overfitting of $f(6,\xi)$ can be found clearly from the coefficient $a_j$ by noticing that $|a_4|>|a_3|$.

\begin{figure}[htbp]
\begin{center}
 \includegraphics[width=0.7\textwidth]{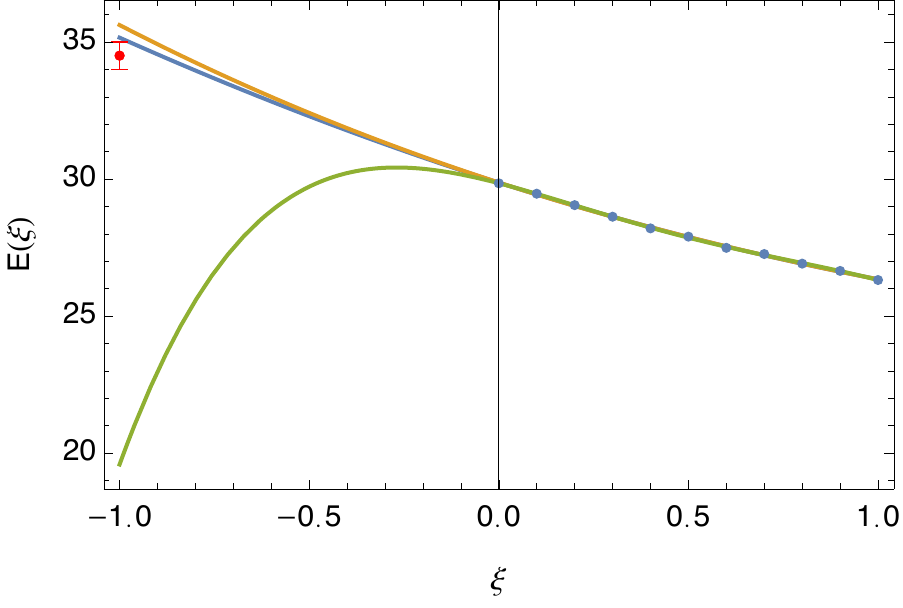} 
\caption{\label{8int} The fitting for $N=8$, $\beta=1$ with Coulomb-type interaction $\lambda=0.5$. We see that $f(2,\xi)$ (blue line) and $f(3,\xi)$ (orange line) can give reliable result, while $f(4,\xi)$ (green line) is an overfitting. The red circle with error bar is the traditional result with path integral Monte Carlo simulations \cite{Dornheim}. $\left<s\right>_B=0.0043$ in the traditional method \cite{Dornheim}, while $\left<s\right>_B\sim 0.002$ based on the estimation of Eq. (\ref{Sequation}).}
\end{center}
\end{figure}

Now we turn to consider more particles. To verify our method, we first consider several examples which have existing results of the energy for fermions by the traditional method \cite{Dornheim}.

\textbf{Example 1:} $N=8$ and $\beta=1$ with Coulomb-type interaction $\lambda=0.5$.

The fitting and the data by PIMD are shown in Fig. \ref{8int}. In addition, the fitting functions are
\begin{equation}
f(2,\xi)=29.8654-4.40772\xi+0.883901\xi^2,
\end{equation}
\begin{equation}
f(3,\xi)=29.8708-4.49251\xi+1.10624\xi^2-0.148226\xi^3,
\end{equation}
\begin{equation}
f(4,\xi)=29.8451-3.60294\xi-3.3416\xi^2+6.96832\xi^3-3.55828\xi^3.
\end{equation}
We see that $f(4,\xi)$ is an obvious overfitting.

\textbf{Example 2:} $N=10$ and $\beta=1$ with Coulomb-type interaction $\lambda=0.5$.

\begin{figure}[htbp]
\begin{center}
 \includegraphics[width=0.7\textwidth]{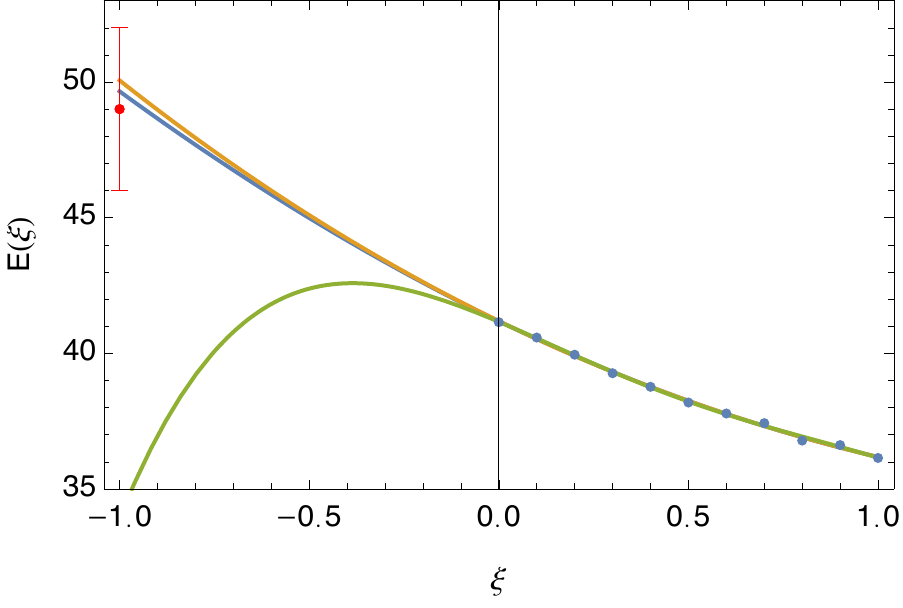} 
\caption{\label{10int} The fitting for $N=10$, $\beta=1$ with Coulomb-type interaction $\lambda=0.5$. The red circle with large error bar is the result of the traditional method \cite{Dornheim}. The blue line is the fitting with $f(2,\xi)$. The orange line is the fitting with $f(3,\xi)$, while the green line is the fitting with $f(4,\xi)$ which shows an obvious overfitting. $\left<s\right>_B=0.0003$ in the traditional method, while $\left<s\right>_B\sim 0.0001$ based on the estimation of Eq. (\ref{Sequation}). We see that in our method, the fluctuation of the energy for fermions is much smaller, compared to the traditional method, although we use much less computing resources.}
\end{center}
\end{figure}

In Fig. \ref{10int}, we give the result for $N=10$ and $\beta=1$ with Coulomb-type interaction $\lambda=0.5$. The red circle with large error bar \cite{Dornheim} is the result of the traditional method to consider the fermion sign problem. 
The fitting functions for the example of Fig. \ref{10int} are
\begin{equation}
f(2,\xi)=41.183-6.73882\xi+1.73446\xi^2,
\end{equation}
\begin{equation}
f(3,\xi)=41.1877-6.81303\xi+1.92907\xi^2-0.12974\xi^3,
\end{equation}
\begin{equation}
f(4,\xi)=41.1619-5.918\xi-2.54609\xi^2+7.03052\xi^3-3.58013 \xi^4.
\end{equation}
In this example, $f(4,\xi)$ is an obvious overfitting.

In the above two examples, the Fermi temperature for 8 particles is $T_F=4$, while it is $T_F=4.49$ for 10 particles. Hence, what we consider is the low temperature situation far below the Fermi temperature. We have checked that for both situations, the extrapolation can not predict the correct value for the energy of fermions without interparticle interaction. When interparticle interaction is considered, however, we notice that the parametrized partition function can give reasonable prediction.

Now we consider large fermion systems which are much beyond the capability of the traditional method.
In Table \ref{energy}, we give the predicted energy of fermions for different particle number. In this table, we also give the uncertainty of the energy for fermions. In the error analysis, we should not underestimate the uncertainty. In all our examples in Table \ref{energy}, we find that both $f(2,\xi)$ and $f(3,\xi)$ can give good fitting for the data of $E(\xi\geq 0)$, while it is possible that $f(n\geq 4,\xi)$ will lead to the overfitting. In this case, we use the average value of the prediction with $f(2,\xi)$ and $f(3,\xi)$ as the energy of fermions, while the difference as the uncertainty. Because the mean deviation for $\xi\geq 0$ is much smaller than this estimation of the uncertainty, the error mainly comes from the extrapolation. In our previous work \cite{XiongFSP}, we estimate that the uncertainty is one order of magnitude of the mean deviation of  the fitting for $\xi\geq 1$. This analysis is consistent with the error analysis in this paper.

\begin{table}[ht]
\caption{Energy for different number of fermions with Coulomb interaction $\lambda=0.5$ and $\beta=1$ in two-dimensional harmonic trap. Here 64.17(19) represents $64.17\pm 0.19$. The same rule applies to other data. Here P-PIMD denotes the method of parametrized path integral molecular dynamics with the data $E(\beta,\xi\geq 0)$ for extrapolation. For $N\leq 10$, the sign factor $s$ is the calculation by Dornheim \cite{Dornheim}, while for $N>10$, it is estimated based on the exponential decreasing of $s$ with the increasing the particle number.} 
\centering 
\begin{tabular}{c c c c} 
\hline\hline 
N~~~ & s & ~~Dornheim\cite{Dornheim}~~ & ~~P-PIMD~~   \\ [0.5ex] 
\hline 
8~~ & 0.00428(6) & 34.5(5) & 34.86(26) \\ 
9~~ & 0.00130(8) & 40(2) & 40.7(4)  \\
10~~ & 0.00030(3) & 49(3) & 49.93(27) \\
11~~ & 0.0001 &  & 55.5(5)  \\
12~~ & 0.00005 &  & 64.17(19)  \\
13~~ & 0.000018 &  & 73.0(8)  \\
14~~ & $6\times 10^{-6}$ &  & 82.8(10)  \\
15~~ & $2\times 10^{-6}$ &  & 88.7(10)  \\
20~~ & $8\times10^{-9}$ &  & 141.5(17) \\ [1ex] 
\hline 
\end{tabular}
\label{energy} 
\end{table}

It is worth pointing out that the main purpose of the present work is to propose and verify the idea of the parametrized PIMD. Hence, we only use moderate $2\times 10^7$ MD steps with separate Nóse-Hoover thermostat \cite{Nose1,Nose2,Hoover,Martyna,Jang}  and $P=12$ in our calculation to test our idea and also show the efficiency of our method, so that the readers can follow our work with relative ease.  
In practical applications, to satisfy the high precision calculation of some problems, we may consider to increase significantly the MD steps, the number of beads $P$ per particle, and also the number of data of $E(\beta,\xi)$ to increase further the accuracy of the energy for fermions. In all the examples of the present work, we have checked the convergence and think that the present MD steps and the number of data are sufficient to give accurate predication, much beyond the traditional method.

Similarly to the case of noninteracting particles at high temperature, the numerical studies of repulsive interaction suggest that for sufficiently strong repulsive interaction, $E(\xi)$ will approach a linear function in the region $-1\leq\xi\leq 1$. In this case, we have
\begin{equation}
E(\xi=-1,\beta)-E(\xi=0,\beta)\simeq E(\xi=0,\beta)-E(\xi=1,\beta).
\end{equation}
Of course, it is not known in advance the validity condition of the above universal relation. However, the accurate numerical calculation for $\xi\geq 0$ can give us the evidence for the validity of the above equation. If $E(\xi\geq 0)$ satisfies well the linear relation, we think the above relation should be satisfied in the whole region of $-1\leq\xi\leq 1$.

\begin{figure}[htbp]
\begin{center}
 \includegraphics[width=0.7\textwidth]{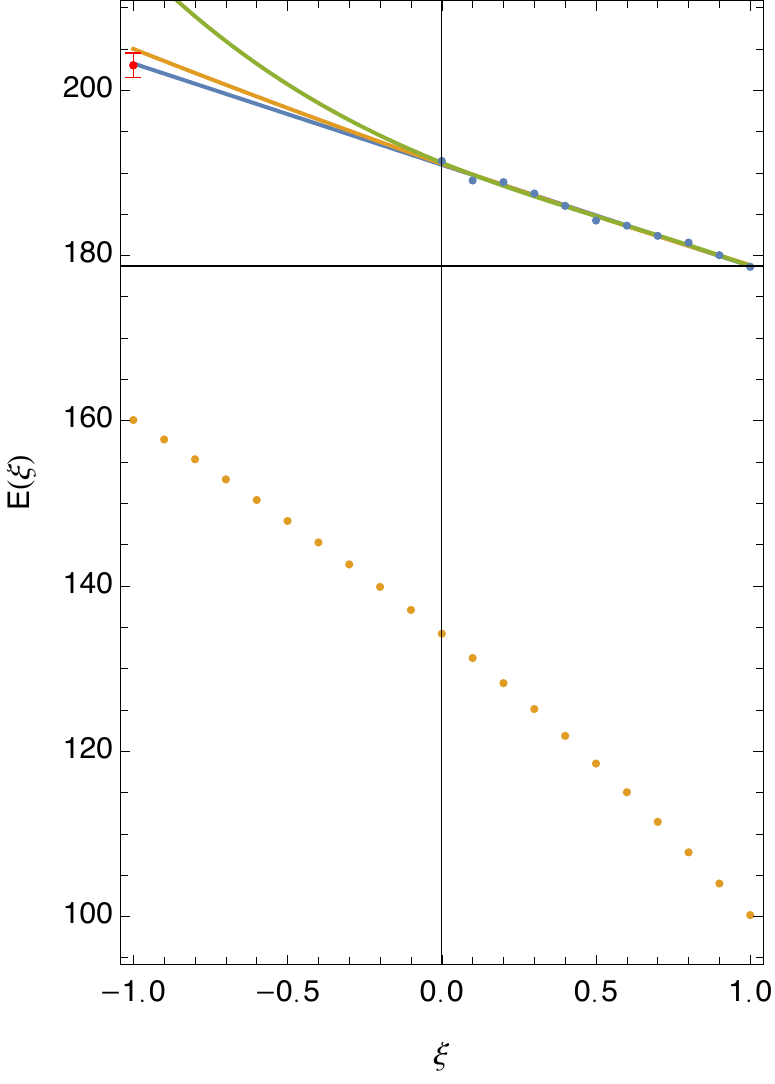} 
\caption{\label{20intstand} For $N=20,\beta=0.3$, and $\lambda=0.5$, we give the result based on different fitting function. Because the data for $E(\xi\geq 0)$ shows a linear behavior, we also use linear function for fitting. The red circle is the result of the traditional method \cite{Dornheim}. We see that both $f(1,\xi)$ (blue line) and $f(2,\xi)$ (orange line) can give good prediction, while $f(3,\xi)$ (green line) is an obvious overfitting. The orange circle is the energy for noninteracting particles. $\left<s\right>_B=0.01$ in the traditional method, while $\left<s\right>_B\sim 0.02$ based on the estimation of Eq. (\ref{Sequation}).}
\end{center}
\end{figure}

To numerically verify the above conjecture, we consider the example of $N=20$ and $\beta=0.3$ with Coulomb-type interaction $\lambda=0.5$ in
Fig. \ref{20intstand}. The blue circle is the result of parametrized PIMD, while the blue line is the fitting with linear function. The red circle with error bar is the result of the traditional method. The linear fit gives the prediction of $E(\xi=-1)=203.2$, while the result of the traditional method is  $E(\xi=-1)=203\pm 1$.  The parabolic function (orange line) also gives a good prediction without suffering from the overfitting, while $f(3,\xi)$ leads to the overfitting because it differs significantly from the consistent result of  $f(1,\xi)$ and $f(2,\xi)$. The fitting functions are
\begin{equation}
f(1,\xi)=190.967-12.2595\xi,
\end{equation}
\begin{equation}
f(2,\xi)=190.09-13.0786\xi+0.819056\xi^2,
\end{equation}
\begin{equation}
f(3,\xi)=191.216-15.0766\xi+6.05857\xi^2-3.49301\xi^3.
\end{equation}
In the same figure, we also show the energy for noninteracting particles (orange circle) with $N=20$ and $\beta=0.3$. We see a clear change of $E(\xi)$ from a concave function for noninteracting particles to a linear function for interacting particles with $\lambda=0.5$.

\section{Perturbation expansion of the parametrized partition function}

After all these examples and the numerical verification of the linear relation, in hindsight, we may understand these results directly from the expression of the parametrized partition function given by Eq. (\ref{Xipartition}). In principle, we may always expand $Z(\beta,\xi)$ as
\begin{equation}
Z(\beta,\xi)=g_0(\beta)+g_1(\beta)\xi+g_2(\beta)\xi^2+\cdots g_{N_p}\xi^{N_p}+\cdots.
\end{equation}
Here the summation is about $N_p$. At first sight, we may think that this expansion expression is useless because we do not know how to calculate all these coefficients $g_j(\beta)$. Because $N_p$ is  the minimum number of times for which pairs of indices must be interchanged in permutation $p$ to recover the original order, for many situations such as sufficiently strong repulsive interaction or sufficiently high temperature, the quantum statistics effect for $g_j(\beta)$ decreases with the increasing of $N_p$. In this situation, we may regard $Z(\beta,\xi)$ as a perturbation expansion about $\xi^{N_p}$. When the condition of this perturbation expansion is satisfied, the energy can be also expanded by the perturbation expansion as follows:
\begin{equation}
E(\beta,\xi)=f_0(\beta)+f_1(\beta)\xi+f_2(\beta)\xi^2+\cdots f_{N_p}\xi^{N_p}+\cdots.
\end{equation}

In all the examples in this paper, what we consider is the determination of the coefficients $f_j(\beta)$ with the energy for $\xi\geq 0$. Of course, the reliable determination of these coefficients relies on whether the condition of the perturbation expansion is valid. In all the examples in this paper, disregarding the case of the overfitting, we notice that $|f_n(\beta)|$ ($n\geq 1$) decreases exponentially as $n$ increases. This provides a clear way to judge whether there exists an overfitting, and also gives the reason why the parabolic function has already given good predication for the energy of fermions. This exponential decay of the coefficient $|f_n(\beta)|$ is not surprising considering the fact that the probability of $N_p$ is proportional to $\alpha ^{N_p}$ and the request of $\alpha<<1$ for the validity condition of our polynomial fitting and extrapolation. This exponential decay of the coefficient $|f_n(\beta)|$ is the physical reason why in Table \ref{energy}, $f(2,n)$ and $f(3,n)$ are sufficient to predict the energy of fermions and give the uncertainty.

It is clear that for the perturbative expansion with exponential decay of the coefficients $|f_n(\beta)|$ ($n\geq 2$), $E(\xi)$ will have good analytical property without the inflection point. The physical analysis of the perturbation expansion about $\xi^{N_p}$ explains why the increasing of the temperature or the increasing of the repulsive interaction contribute to the good analytical property of $E(\xi)$. The so-called linear relationship found in this work is nothing more than the condition that only the terms for  $N_p=0$ and $N_p=1$ are dominant. We emphasize that we may still have significant effect of the quantum statistics because the term with $N_p=1$ may have important physical effect in some situations. When the quantum statistics is negligible, it means that we may approximate $f_1(\beta)$ as zero. The linear relation would be useful to consider the analytical approximation for the energy of fermions, because it is much easier to get the analytical approximation for distinguishable particles and identical bosons. This provides new chance to reconsider unsolved problems for fermions by firstly considering bosons and distinguishable particles. The present work also provides a new perturbative expansion method based on the exact numerical calculation for $\xi\geq 0$, compared with the cluster expansion and the usual perturbative expansion in statistical field theory. One of the advantages of our approach is the connection of the quantum systems with different statistics.

\section{Summary and discussion}

In summary, after systematic studies on noninteracting and repulsively interacting particles, we have the following schematic diagram for the validity of the method of parametrized PIMD. Essentially, we provide an efficient method to consider the numerically perturbative expansion about $\xi^{N_p}$. In future work, we will study the case of attractive interaction. The present work makes it feasible to consider the thermodynamics of large fermion systems with repulsive interaction by circumventing the fermion sign problem. Because we provided reliable calculation results (Table \ref{energy}) for large fermion systems much beyond the capability of the traditional method, our result or method will provide excellent benchmark for other methods trying to overcome the fermion sign problem for large fermion systems. It seems that our method has good chance to expand the valid regions of other methods such as density matrix quantum Monte Carlo \cite{Blunt,Malone}, configuration PIMC \cite{Schoof1, Schoof2,Schoof3, Yilmaz} and permutation blocking PIMC \cite{PB1,PB2} for the simulation of warm dense matter \cite{WDM}. In addition, compared with configuration PIMC and permutation blocking PIMC, the present work is more straightforward to consider nonuniform fermion systems.
It is clear that  the direct calculation of $E(\xi)$ for $\xi\leq 0$ would be useful to overcome the fermion sign problem, even for negative $\xi$ not too far from zero. This is a problem deserving further studies based on the formulation of the parametrized path integral partition function in this work.

\begin{figure}[htbp]
\begin{center}
 \includegraphics[width=0.7\textwidth]{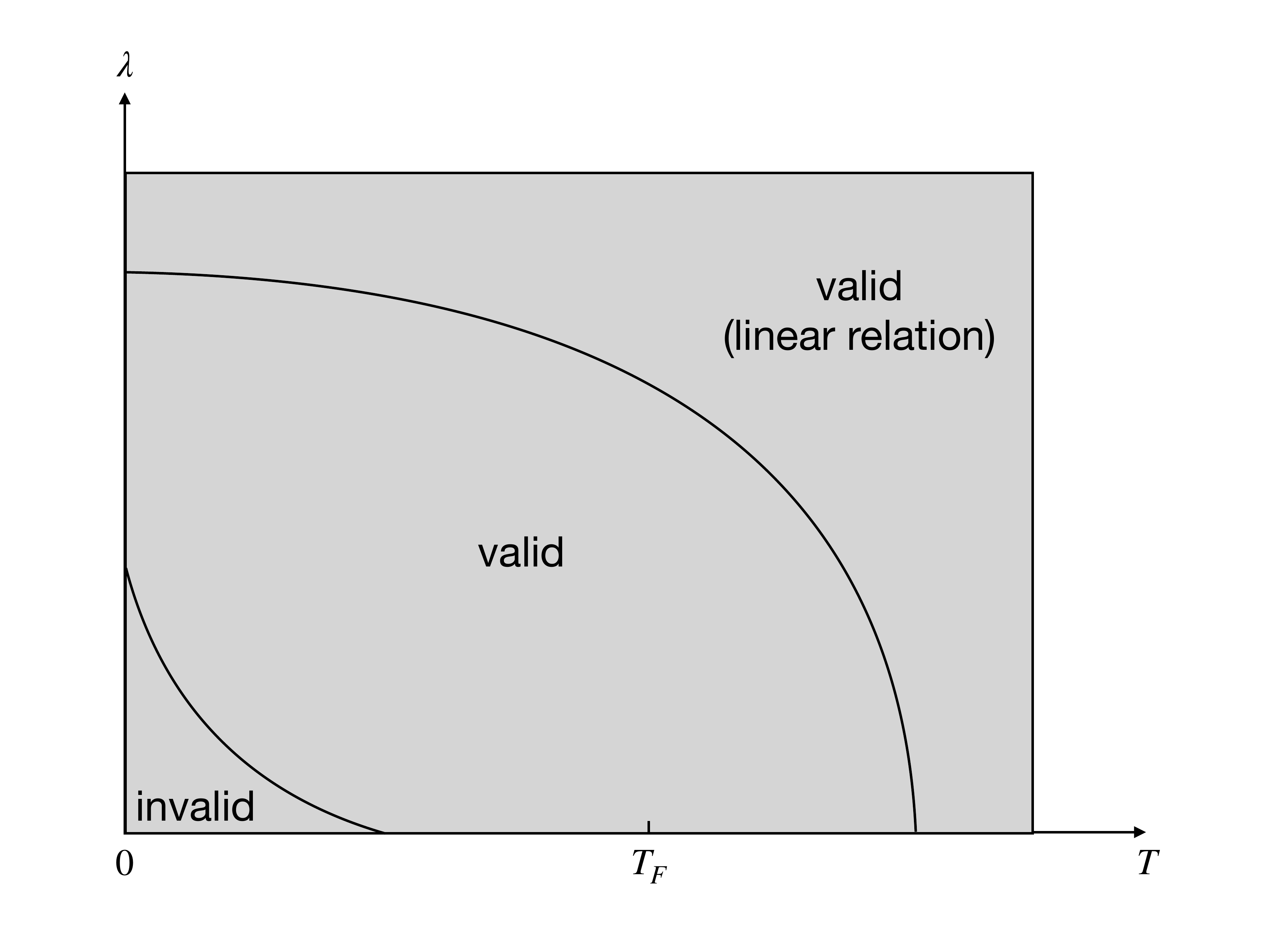} 
\caption{\label{validity} Illustration of the invalid region, valid region and valid region satisfying the linear relation.}
\end{center}
\end{figure}

\begin{acknowledgments}
This work is partly supported by the National Natural Science Foundation of China under grant numbers 11175246, and 11334001. 
\end{acknowledgments}

\textbf{DATA AVAILABILITY}

The data that support the findings of this study are available from the corresponding author upon reasonable request. The code of this study is openly available in GitHub (https://github.com/xiongyunuo/PIMD-Pro).


\begin{thebibliography}{10}


\bibitem{Hirshberg} B. Hirshberg, V. Rizzi, and M. Parrinello, Path integral molecular dynamics for bosons, 
\text{Proc. Natl. Acad. Sci. U. S. A.}~\textbf{116}, 21445 (2019).

\bibitem{Deuterium}   C. W. Myung, B. Hirshberg, and M. Parrinello, 
Prediction of a supersolid phase in high-pressure deuterium, 
\text{Phys. Rev. Lett.} \textbf{128}, 045301 (2022).

\bibitem{CeperRMP} D. M. Ceperley, 
Path integrals in the theory of condensed helium, 
\text{Rev. Mod. Phys.} \textbf{67}, 279 (1995).

\bibitem{boninsegni1} M.~Boninsegni, N. V.~Prokof’ev, and B. V.~Svistunov, 
Worm algorithm and diagrammatic Monte Carlo: A new approach to continuous-space path integral Monte Carlo simulations, 
 {\text{Phys.~Rev.~E}~\textbf{74}, 036701} (2006).

\bibitem{boninsegni2} M.~Boninsegni, N. V.~Prokof’ev, and B. V.~Svistunov, 
Worm algorithm for continuous-space path integral Monte Carlo simulations,  
{\text{Phys.~Rev.~Lett.}~\textbf{96}, 070601} (2006).

\bibitem{Dornheim} T.~Dornheim, 
The Fermion sign problem in path integral Monte Carlo simulations: quantum dots, ultracold atoms, and warm dense matter, 
\text{Phys. Rev. E}~\textbf{100}, 023307 (2019).

\bibitem{DornheimMod} T. Dornheim, M. Invernizzi, J. Vorberger, and B. Hirshber, 
Attenuating the fermion sign problem in path integral Monte Carlo simulations using the Bogoliubov inequality and thermodynamic integration, 
J. Chem. Phys. \textbf{153}, 234104 (2020).

\bibitem{ceperley} D. M.~Ceperley, Path Integral Monte Carlo Methods for Fermions, Monte Carlo and Molecular Dynamics of Condensed Matter Systems, K.~Binder and G.~Ciccotti (Eds.), Editrice Compositori, Bologna (Italy) (1996).

\bibitem{Alex} A. Alexandru, G. Basar, P. F. Bedaque, and N. C. Warrington, 
Complex paths around the sign problem, 
Rev. Mod. Phys. \textbf{94}, 015006 (2022).

\bibitem{troyer} M.~Troyer and U. J.~Wiese, 
Computational Complexity and Fundamental Limitations to Fermionic Quantum Monte Carlo Simulations, 
{\text{Phys. Rev. Lett.} \textbf{94}, 170201} (2005).


\bibitem{loh} E. Y.~Loh, J. E.~Gubernatis, R. T.~Scalettar, S. R.~White, D. J.~Scalapino, and R. L.~Sugar, 
Sign problem in the numerical simulation of many-electron systems, 
{\text{Phys. Rev. B} \textbf{41}, 9301} (1990).


\bibitem{lyubartsev} A. P.~Lyubartsev, 
Simulation of excited states and the sign problem in the path integral Monte Carlo method, 
{\text{J.~Phys.~A: Math.~Gen.}~\textbf{38}, 6659} (2005).

\bibitem{vozn} M. A.~Voznesenskiy, P. N.~Vorontsov-Velyaminov, and A. P.~Lyubartsev, 
Path-integral-expanded-ensemble Monte Carlo method in treatment of the sign problem for fermions, 
\text{Phys.~Rev.~E}~\textbf{80}, 066702 (2009).


\bibitem{Science} R. Mondaini, S. Tarat, and R. T. Scalettar,  Quantum critical points and the sign problem, Science \textbf{375}, 
418 (2022).


\bibitem{Wu} Congjun Wu and Shou-Cheng Zhang, 
Sufficient condition for absence of the sign problem in the fermionic quantum Monte Carlo algorithm, 
Phys. Rev. B \textbf{71}, 155115 (2005).

\bibitem{Umrigar} C. J. Umrigar, Julien Toulouse, Claudia Filippi, S. Sorella, and R. G. Hennig, 
Alleviation of the Fermion-Sign Problem by Optimization of Many-Body Wave Functions, 
Phys. Rev. Lett. \textbf{98}, 110201 (2007).

\bibitem{Li} Zi-Xiang Li, Yi-Fan Jiang, and Hong Yao, 
Solving the fermion sign problem in quantum Monte Carlo simulations by Majorana representation, 
Phys. Rev. B \textbf{91}, 241117(R) (2015).

\bibitem{Wei} Z. C. Wei, Congjun Wu, Yi Li, Shiwei Zhang, and T. Xiang, 
Majorana Positivity and the Fermion Sign Problem of Quantum Monte Carlo Simulations, 
Phys. Rev. Lett. \textbf{116}, 250601 (2016).

\bibitem{Yao2} Z. X.~Li, Y.~F. Jiang, and H.~Yao, Majorana-Time-Reversal Symmetries: A Fundamental Principle for Sign-Problem-Free Quantum Monte Carlo Simulations, { \text{Phys.~Rev.~Lett.}~\textbf{117}, 267002} (2016).

\bibitem{XiongFSP} Y. Xiong and H. Xiong, On the thermodynamic properties of fictitious identical particles and the application to fermion sign problem, J. Chem. Phys. (2022). https://doi.org/10.1063/5.0106067  


\bibitem{HirshbergFermi} B. Hirshberg,  M. Invernizzi, and  M. Parrinello, 
Path integral molecular dynamics for fermions: Alleviating the sign problem with the Bogoliubov inequality, 
\text{J. Chem. Phys.} \textbf{152}, 171102 (2020).


\bibitem{Xiong} Y. Xiong and H. Xiong, 
Path integral molecular dynamics simulations for Green's function in a system of identical bosons, 
J. Chem. Phys. \textbf{156}, 134112 (2022).

\bibitem{Xiong2} Y. Xiong and  H. Xiong, 
Numerical calculation of Green's function and momentum distribution for spin-polarized fermions by path integral molecular dynamics, 
J. Chem. Phys. \textbf{156}, 204117 (2022).

\bibitem{Xiong3} Y. Xiong and  H. Xiong, Path integral and winding number in singular magnetic field, Eur. Phys. J. Plus \textbf{137}, 550 (2022).

\bibitem{Xiong4} Y. Yu, S. Liu, H. Xiong, and Y. Xiong, Path integral molecular dynamics for thermodynamics and Green's function of ultracold spinor bosons, J. Chem. Phys. \textbf{157}, 064110 (2022).

\bibitem{Xiong5} Y. Xiong and H. Xiong, Path integral molecular dynamics for anyons, bosons and fermions, Phys. Rev. E 106, 025309 (2022).


\bibitem{nodes} D. M. Ceperley, Fermion nodes, J. Stat. Phys., \textbf{63}, 1237 (1991).

\bibitem{Helium}  D. M. Ceperley, Path-integral calculations of normal liquid 
3He, Phys. Rev. Lett. \textbf{69}, 331 (1992).

\bibitem{Militzer} B. Militzer, E. L. Pollock, and D. M. Ceperley, Path integral Monte Carlo calculation of the momentum distribution of the homogeneous electron gas at finite temperature, High Energy Dens. Phys. \textbf{30}, 13 (2019).


\bibitem{Blunt} N. S. Blunt, T. W. Rogers, J. S.  Spencer, and W. M. Foulkes,
Density-matrix quantum Monte Carlo method, Phys.
Rev. B \textbf{89}, 245124 (2014). 

\bibitem{Malone} F.D. Malone, N. S. Blunt, James J. Shepherd, D. K. K. Lee,  J. S. Spencer, and  W. M. C. Foulkes, Interaction Picture Density Matrix Quantum Monte Carlo, J. Chem. Phys. \textbf{143}, 044116 (2015).


\bibitem{Schoof1} T. Schoof, M. Bonitz, A. V. Filinov, D. Hochstuhl and
J. W. Dufty, Configuration Path Integral Monte Carlo,
Contrib. Plasma Phys. \textbf{51}, 687 (2011).

\bibitem{Schoof2} T. Schoof, S. Groth,  and M. Bonitz, Towards ab Initio Thermodynamics of the Electron Gas at Strong Degeneracy, Contrib. Plasma Phys. \textbf{55}, 136 (2015).

\bibitem{Schoof3} T. Schoof, S. Groth, J. Vorberger,  and M. Bonitz, Ab Initio Thermodynamic Results for the Degenerate Electron Gas at Finite Temperature, Phys. Rev. Lett. \textbf{115}, 130402 (2015).

\bibitem{Yilmaz} A. Yilmaz, K. Hunger,  T. Dornheim, S. Groth, and  M. Bonitz, Restricted configuration path integral Monte Carlo, J. Chem. Phys. \textbf{153}, 124114 (2020).

\bibitem{PB1} T. Dornheim, S. Groth, A. Filinov, and M. Bonitz, Permutation blocking path integral Monte Carlo: a highly efficient approach to the simulation of strongly degenerate non-ideal fermions, New J. Phys. \textbf{17}, 073017 (2015).

\bibitem{PB2} T. Dornheim, T. Schoof, S. Groth, A. Filinov, and M. Bonitz, Permutation blocking path integral Monte Carlo approach to the uniform electron gas at finite temperature, J. Chem. Phys. \textbf{143},  204101 (2015).


\bibitem{Joonho} Joonho Lee, Miguel A. Morales, and Fionn D. Malone, A phaseless auxiliary-field quantum Monte Carlo perspective on the uniform electron gas at finite temperatures: Issues, observations, and benchmark study, J. Chem. Phys. \textbf{154}, 064109 (2021).




\bibitem{feynman} R. P.~Feynman and A. R.~Hibbs, Quantum mechanics and path integrals, Dover Publications, New York (2010).

\bibitem{kleinert} H.~Kleinert, Path integrals in quantum mechanics, statistics, polymer physics, and financial markets, World Scientific, Singapore (2009).

\bibitem{Tuckerman} M. E.~Tuckerman, Statistical mechanics: theory and molecular simulation, Oxford University, New York (2010).


 \bibitem{chandler} D.~Chandler and P. G.~Wolynes, Exploiting the isomorphism between quantum theory and classical statistical mechanics of polyatomic fluids, {\text{J.~Chem.~Phys.}~\textbf{74}, 4078} (1981).

\bibitem{Parrinello} M.~Parrinello and A.~Rahman, Study of an F center in molten KCl, \text{J. Chem. Phys.}~\textbf{80}, 860 (1984).

\bibitem{Miura} S. Miura and S. Okazaki, Path integral molecular dynamics for Bose-Einstein and Fermi-Dirac statistics. \text{J. Chem. Phys.} \textbf{112}, 10116 (2000).

 \bibitem{Cao} J. Cao and G. A. Voth,  The formulation of quantum statistical mechanics based on the Feynman path centroid density. I. Equilibrium properties, \text{J. Chem. Phys.} \textbf{100},  5093 (1994).
 
 \bibitem{Cao2} J. Cao and G. A. Voth, The formulation of quantum statistical mechanics based on the Feynman path centroid density. II. Dynamical properties,  \text{J. Chem. Phys.} \textbf{100}, 5106 (1994).
 
 \bibitem{Jang2} S. Jang and G. A. Voth, A derivation of centroid molecular dynamics and other approximate time evolution methods for path integral centroid variables,  \text{J. Chem. Phys.} \textbf{111}, 2371 (1999).
 
\bibitem{Ram} R. Ram\'iRez and T. L\'oPez-Ciudad, The Schr\"odinger formulation of the Feynman path centroid density, \text{J. Chem. Phys.} \textbf{111}, 3339 (1999).


\bibitem{Kinugawa} K. Kinugawa, H. Nagao, and K. Ohta, Path integral centroid molecular dynamics method for Bose and Fermi statistics: formalism and simulation, Chem. Phys. Lett. \textbf{307}, 187 (1999).

\bibitem{Roy} Pierre-Nicholas Roy, Seogjoo Jang, and Gregory A. Voth, Feynman path centroid dynamics for Fermi–Dirac statistics, J. Chem. Phys. \textbf{111}, 5303 (1999).

\bibitem{Roy1} P.-N. Roy and G.A. Voth, On the Feynman path centroid density for Bose-Einstein and Fermi-Dirac statistics, J. Chem. Phys. \textbf{110}, 3647 (1999).


\bibitem{Roy2} N. Blinov, P.-N. Roy, and G.A. Voth, Path integral formulation of centroid dynamics for systems obeying Bose-Einstein statistics, J. Chem. Phys. \textbf{115}, 4484 (2001).

\bibitem{Roy3} N. Blinov and P.-N. Roy, Operator formulation of centroid dynamics for Bose-Einstein and Fermi-Dirac statistics, J. Chem. Phys. \textbf{115}, 7822 (2001).

\bibitem{Kinugawa1} K. Kinugawa, H. Nagao, and K. Ohta, A semiclassical approach to the dynamics of many-body Bose/Fermi systems by the path integral centroid molecular dynamics, J. Chem. Phys. \textbf{114}, 1454 (2001).

\bibitem{Kinugawa2} K. Kinugawa, H. Nagao, and K. Ohta, A path integral centroid molecular dynamics method for Bose and Fermi statistics, J. Mol. Liq. \textbf{90}, 11 (2001).

\bibitem{Roy4} Nicholas Blinov and Pierre-Nicholas Roy, An effective centroid Hamiltonian and its associated centroid dynamics for indistinguishable particles in a harmonic trap, J. Chem. Phys. \textbf{116}, 4808 (2002).

\bibitem{Roy5} Pierre-Nicholas Roy and Nicholas Blinov, Centroid dynamics with quantum statistics, Isr J. Chem. \textbf{42}, 183 (2002).

\bibitem{Roy6} Paul Moffatt, Nicholas Blinov, and 
Pierre-Nicholas Roy, On the calculation of single-particle time correlation functions from Bose–Einstein centroid dynamics, J. Chem. Phys. \textbf{120}, 4614 (2004).

 
\bibitem{Poly} E. A. Polyakov, A. P.  Lyubartsev, and P. N.  Vorontsov-Velyaminov,  Centroid molecular dynamics: Comparison with exact results for model systems, \text{J. Chem. Phys.} \textbf{133}, 194103 (2010). 

 \bibitem{Craig} I. R. Craig and D. E. Manolopoulos, Quantum statistics and classical mechanics: Real time correlation functions from ring polymer molecular dynamics,  \text{J. Chem. Phys.}  \textbf{121}, 3368 (2004). 
 
 \bibitem{Braa} B. J. Braams and D. E. Manolopoulos, On the short-time limit of ring polymer molecular dynamics, \text{J. Chem. Phys.} \textbf{125}, 124105 (2006).
 
 \bibitem{Haber} S. Habershon, D. E. Manolopoulos, T. E. Markland, and T. F. Miller 3rd, Ring-polymer molecular dynamics: quantum effects in chemical dynamics from classical trajectories in an extended phase space, \text{Annu. Rev. Phys. Chem.} \textbf{64}, 387 (2013).
 
 \bibitem{Thomas} T. E. Markland and M. Ceriotti, Nuclear quantum effects enter the mainstream, \text{Nat. Rev. Chem.} \textbf{2}, 0109 (2018).
 








\bibitem{Nose1} S. Nos\'e, 
A molecular dynamics method for simulations in the canonical ensemble, 
\text{Mol. Phys.} \textbf{52}, 255 (1984).

\bibitem{Nose2} S. Nos\'e, 
A unified formulation of the constant temperature molecular dynamics methods, 
\text{J. Chem. Phys.} \textbf{81}, 511 (1984).

\bibitem{Hoover} W. G. Hoover, 
Canonical dynamics: Equilibrium phase-space distributions, 
\text{Phys. Rev. A} \textbf{31}, 1695 (1985).

\bibitem{Martyna} G. J. Martyna, M. L. Klein, and M. Tuckerman, 
Nos\'e-Hoover chains: The canonical ensemble via continuous dynamics, 
\text{J. Chem. Phys.} \textbf{97}, 2635 (1992).

\bibitem{Jang} S. Jang and G. A. Voth, 
Simple reversible molecular dynamics algorithms for Nos\'e-Hoover chain dynamics, 
\text{J. Chem. Phys.}~\textbf{107}, 9514 (1997).



\bibitem{WDM} T. Dornheim, S. Groth, and M. Bonitz, The uniform electron gas at warm dense matter conditions, Phys. Rep \textbf{744}, 1
(2018).



\end{thebibliography}
\end{document}